\documentclass[journal]{IEEEtran}

\usepackage{cite}
\usepackage{amsmath,amssymb,amsfonts}
\usepackage{graphicx}
\usepackage{textcomp}
\usepackage{xcolor}
\usepackage{booktabs}
\usepackage{multirow}
\usepackage{tikz}
\usetikzlibrary{positioning,arrows.meta,calc,fit,backgrounds,decorations.pathreplacing}
\usepackage{url}
\usepackage{balance}

\def\BibTeX{{\rm B\kern-.05em{\sc i\kern-.025em b}\kern-.08em
    T\kern-.1667em\lower.7ex\hbox{E}\kern-.125emX}}

\begin{document}

\title{From 8 Seconds to 370\,ms: Kernel-Fused SAR Imaging\\on Apple Silicon via Single-Dispatch FFT Pipelines}

\author{\IEEEauthorblockN{Mohamed Amine Bergach}\\
\IEEEauthorblockA{Illumina\\
mbergach@illumina.com}
\thanks{Source code: \protect\url{https://github.com/aminems/AppleSiliconFFT}}}

%\markboth{IEEE Signal Processing Letters}%
%{Bergach: From 8 Seconds to 370\,ms: Kernel-Fused SAR Imaging on Apple Silicon}

\maketitle

% =============================================================================
% ABSTRACT
% =============================================================================
\begin{abstract}
We present the first kernel-fused SAR Range Doppler pipeline on any GPU
platform. By fusing FFT, matched-filter multiply, and IFFT into a single Metal
compute dispatch---keeping all intermediate data in 32\,KiB on-chip
memory---we process a $4096\!\times\!4096$ complex SAR scene in
\textbf{370\,ms} on an Apple M1 GPU, a \textbf{22$\times$} speedup over
the multi-dispatch baseline (8.16\,s). We further report the first FFT
to exploit Apple's \texttt{simdgroup\_matrix} 8$\times$8 hardware MMA,
enabled by an in-place Cooley--Tukey decimation-in-frequency formulation that
halves the memory footprint versus Stockham. Radar image quality is
preserved: all five point targets show 0.0\,dB SNR deviation from the
unfused FP32 reference.
\end{abstract}


% =============================================================================
% I. INTRODUCTION
% =============================================================================
\section{Introduction}

\IEEEPARstart{S}{ynthetic} Aperture Radar (SAR) imaging relies on compute-intensive
frequency-domain operations---primarily batched FFT, matched-filter multiplication,
and IFFT---repeated across thousands of range and azimuth
lines~\cite{cumming2005sar,moreira2013sar_tutorial}. GPU acceleration of SAR has
been studied extensively on NVIDIA
hardware~\cite{calore2020sar,zhang2021multi_gpu_bpa,chen2024embedded_visar}, with
speedups of 40--270$\times$ over CPU. However, all published GPU SAR
implementations dispatch FFT, multiply, and IFFT as \emph{separate} kernel
launches~\cite{esser2020cudarange}, incurring redundant device-memory traffic
between stages. Moreover, no SAR implementation exists for Metal or Apple Silicon.

Kernel fusion---combining multiple operations into a single GPU dispatch so that
intermediate data stays in on-chip memory---has been demonstrated for general
signal processing~\cite{nvidia_cufftdx,adamek2020smfft} and neural
operators~\cite{wu2025turbofno}, but never for SAR. The key enablers on NVIDIA
are cuFFTDx (device-side FFT since 2020)~\cite{nvidia_cufftdx} and custom shared-memory
FFT kernels~\cite{adamek2020smfft,wu2025turbofno}. No equivalent exists on
Metal, where the only cross-platform FFT library
(VkFFT~\cite{tolmachev2023vkfft}) does not support fusion.

A companion paper~\cite{bergach2026fft} established a two-tier memory model for
FFT on Apple Silicon: 208~KiB registers (Tier~1) and 32~KiB threadgroup memory
(Tier~2), achieving 138~GFLOPS with a radix-8 Stockham kernel at $N\!=\!4096$.
This paper builds on that foundation with three contributions:

\begin{enumerate}
\item \textbf{First kernel-fused SAR pipeline on any GPU}: FFT $\to$
matched-filter multiply $\to$ IFFT in a single Metal dispatch, with data
resident in 32~KiB threadgroup memory throughout. A 22$\times$ end-to-end
speedup over the unfused baseline on Apple M1.

\item \textbf{First MMA-based FFT on Apple Silicon}: An in-place Cooley-Tukey
decimation-in-frequency (DIF) kernel using \texttt{simdgroup\_matrix}
8$\times$8 MMA for the radix-8 DFT butterfly, achieving 128~GFLOPS (93\% of
the scalar baseline).

\item \textbf{First SAR implementation on Metal/Apple GPU}: Complete Range
Doppler Algorithm with validated point-target quality (PSLR, ISLR, SNR).
\end{enumerate}

% =============================================================================
% II. SYSTEM ARCHITECTURE
% =============================================================================
\section{System Architecture}

\subsection{Apple Silicon Memory Hierarchy}

Apple Silicon GPUs provide a two-tier on-chip storage hierarchy
(detailed in~\cite{bergach2026fft}): a 208~KiB register file (private per
thread, exchangeable within 32-thread SIMD groups via \texttt{simd\_shuffle})
and 32~KiB threadgroup memory shared across all threads. Device memory is
unified with the CPU---zero-copy access with hardware coherence.

For $N\!=\!4096$ complex float32, the working set is exactly 32~KiB
($4096 \times 8$~bytes), filling the entire threadgroup memory. This makes
$N\!=\!4096$ the largest FFT computable in a single threadgroup without
device-memory exchange---and is the typical range-line length in medium-resolution
SAR systems~\cite{cumming2005sar}.

\subsection{Kernel Fusion Design}

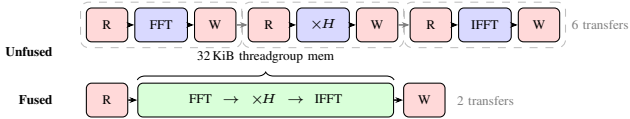
\begin{figure}[t]
\centering
\begin{tikzpicture}[
  box/.style={draw, rounded corners=2pt, minimum height=0.45cm,
              font=\tiny, inner sep=2pt},
  mem/.style={box, fill=red!15, minimum width=0.55cm},
  kern/.style={box, fill=blue!15, minimum width=0.7cm},
  fused/.style={box, fill=green!15},
  arr/.style={-{Stealth[length=3pt]}, semithick},
  lbl/.style={font=\tiny\bfseries, anchor=east},
]
% --- Unfused pipeline (two rows) ---
\node[lbl] at (-0.1, 0) {Unfused};

% Row 1: FFT dispatch
\node[mem]  (r1) at (0.5, 0.35) {R};
\node[kern, right=0.08cm of r1] (fft1) {FFT};
\node[mem, right=0.08cm of fft1] (w1) {W};
% Row 1: Multiply dispatch
\node[mem, right=0.15cm of w1]  (r2) {R};
\node[kern, right=0.08cm of r2] (mul1) {$\times\!H$};
\node[mem, right=0.08cm of mul1] (w2) {W};
% Row 1: IFFT dispatch
\node[mem, right=0.15cm of w2]  (r3) {R};
\node[kern, right=0.08cm of r3] (ifft1) {IFFT};
\node[mem, right=0.08cm of ifft1] (w3) {W};

\draw[arr] (r1) -- (fft1);
\draw[arr] (fft1) -- (w1);
\draw[arr,densely dashed,gray] (w1) -- (r2);
\draw[arr] (r2) -- (mul1);
\draw[arr] (mul1) -- (w2);
\draw[arr,densely dashed,gray] (w2) -- (r3);
\draw[arr] (r3) -- (ifft1);
\draw[arr] (ifft1) -- (w3);

% Count label
\node[font=\tiny,gray,anchor=west] at (w3.east) {\,6 transfers};

% --- Fused pipeline ---
\node[lbl] at (-0.1, -0.65) {Fused};
\node[mem]  (fr) at (0.5, -0.65) {R};
\node[fused, right=0.12cm of fr, minimum width=3.4cm] (fk)
  {FFT $\;\to\;$ $\times\!H$ $\;\to\;$ IFFT};
\node[mem, right=0.12cm of fk]  (fw) {W};

\draw[arr] (fr) -- (fk);
\draw[arr] (fk) -- (fw);

% Count label
\node[font=\tiny,gray,anchor=west] at (fw.east) {\,2 transfers};

% Brace for threadgroup memory
\draw[decorate, decoration={brace, amplitude=3pt, raise=1pt}]
  (fk.north west) -- (fk.north east)
  node[midway, above=4pt, font=\tiny] {32\,KiB threadgroup mem};

% Dashed box around each dispatch in unfused
\begin{scope}[on background layer]
  \node[draw=gray!50, dashed, rounded corners=3pt, inner sep=2pt,
        fit=(r1)(w1)] {};
  \node[draw=gray!50, dashed, rounded corners=3pt, inner sep=2pt,
        fit=(r2)(w2)] {};
  \node[draw=gray!50, dashed, rounded corners=3pt, inner sep=2pt,
        fit=(r3)(w3)] {};
\end{scope}

\end{tikzpicture}
\caption{Unfused (top) vs.\ fused (bottom) range compression.
Red: device-memory I/O; blue: compute kernels; green: fused kernel.
Dashed boxes are separate GPU dispatches. Dashed arrows are redundant
device-memory round-trips eliminated by fusion.}
\label{fig:pipeline}
\end{figure}

The unfused SAR compression pipeline (Fig.~\ref{fig:pipeline}, top) requires
three separate dispatches per line: FFT, multiply, IFFT. Each dispatch reads from
and writes to device memory, producing 6 device-memory transfers (3 reads +
3 writes) per range or azimuth line.

Our fused kernel (Fig.~\ref{fig:pipeline}, bottom) combines all three operations
into a single dispatch. Each threadgroup processes one line:

\begin{enumerate}
\item \textbf{Load}: Read $N$ complex samples from device memory into the
32~KiB threadgroup buffer.
\item \textbf{Forward FFT}: Six radix-4 Stockham passes entirely in threadgroup
memory (1024 threads, 4 elements/thread).
\item \textbf{Multiply}: Point-wise complex multiply with the pre-computed
matched filter (loaded from a separate device buffer).
\item \textbf{IFFT}: Conjugate in-place, reuse the same forward FFT passes,
final conjugate + $1/N$ scale on write to device memory.
\item \textbf{Store}: Write $N$ compressed samples to device memory.
\end{enumerate}

Data transfers drop from 6 to 2 (one read, one write). The matched-filter
read (step~3) hits the System Level Cache (SLC) because the same $N$-element
filter is reused by every threadgroup.

\subsection{IFFT via Conjugate-FFT-Conjugate}

The inverse FFT is computed as
$\text{IFFT}(x) = \frac{1}{N}\,\overline{\text{FFT}(\bar{x})}$,
where $\bar{\cdot}$ denotes complex conjugation. This reuses the forward FFT
butterfly \emph{unchanged}, requiring only two additional threadgroup-memory
passes for conjugation (negating the imaginary component). The final conjugation
and $1/N$ scaling are folded into the last Stockham pass's device-memory store,
adding zero extra cost.

% =============================================================================
% III. IN-PLACE CT MMA FFT
% =============================================================================
\section{In-Place Cooley-Tukey MMA FFT}

\subsection{Motivation: Stockham vs.\ Cooley-Tukey for MMA}

The \texttt{simdgroup\_matrix} API exposes 8$\times$8 hardware
MMA~\cite{apple_msl_spec,thundermittens2024}, achieving $\sim$4$\times$ higher
ALU utilization than scalar SIMD~\cite{turner_metal_benchmarks}. The radix-8
DFT maps naturally to an 8$\times$8 complex matrix multiply decomposed into
four real MMA operations~\cite{li2021tcfft}:
\begin{align}
Y_{\text{re}} &= F_{8,\text{re}} \cdot X_{\text{re}} - F_{8,\text{im}} \cdot X_{\text{im}} \\
Y_{\text{im}} &= F_{8,\text{re}} \cdot X_{\text{im}} + F_{8,\text{im}} \cdot X_{\text{re}}
\end{align}

However, MMA requires a \emph{split real/imaginary} layout (separate
\texttt{float} buffers for real and imaginary parts) rather than the interleaved
\texttt{float2} layout used by Stockham. On Apple GPU, each component buffer
requires 16~KiB ($4096 \times 4$~bytes), totaling 32~KiB---exactly filling
threadgroup memory with no room for double-buffering.

The out-of-place Stockham formulation requires two buffers (source and
destination), which would need 64~KiB in split layout---exceeding the 32~KiB
limit. We therefore adopt an \textbf{in-place Cooley-Tukey DIF} formulation
that overwrites input data in place, requiring only a single 32~KiB buffer pair.

\subsection{Kernel Architecture}

The MMA kernel (\texttt{fft\_4096\_ct\_mma}) decomposes the $N\!=\!4096$ FFT
as four radix-8 DIF stages ($4096 = 8^4$) with strides $S \in \{512, 64, 8, 1\}$.
Stages 0--2 use MMA; stage~3 uses scalar butterfly (stride 1, contiguous
elements, no MMA benefit):

\begin{itemize}
\item \textbf{512 threads}, 32 SIMD groups, 4 MMA tiles per SIMD group.
\item \textbf{DFT$_8$ matrix} loaded once into \texttt{simdgroup\_float8x8}
registers and reused across all stages.
\item \textbf{Twiddle application} via \texttt{thread\_elements()}: we
empirically determined the mapping between SIMD lane ID and matrix element
position on Apple M1 (undocumented):
\begin{align}
\text{row} &= \lfloor\text{lane}/16\rfloor \cdot 4 + \lfloor(\text{lane} \bmod 8)/2\rfloor \\
\text{col}_0 &= (\lfloor\text{lane}/8\rfloor \bmod 2) \cdot 4 + (\text{lane} \bmod 2) \cdot 2
\end{align}
Element 0 resides at $(\text{row}, \text{col}_0)$ and element 1 at
$(\text{row}, \text{col}_0\!+\!1)$.
\end{itemize}

The final stage fuses the scalar radix-8 butterfly with digit-reversal
permutation and device-memory output, eliminating an extra barrier and
threadgroup write.

\subsection{Performance Analysis}

Table~\ref{tab:mma} compares the MMA and scalar kernels.

\begin{table}[t]
\centering
\caption{MMA vs.\ Scalar FFT ($N\!=\!4096$, Batch 256, Apple M1)}
\label{tab:mma}
\begin{tabular}{@{}lrrr@{}}
\toprule
\textbf{Kernel} & \textbf{GFLOPS} & \textbf{$\mu$s/FFT} & \textbf{Notes} \\
\midrule
Radix-8 Stockham (scalar) & 138 & 1.78 & Best overall \\
CT MMA (stages 0--2) & 128 & 1.92 & 93\% of scalar \\
CT scalar reference & 120 & 2.04 & Split layout only \\
\bottomrule
\end{tabular}
\end{table}

The MMA kernel achieves 128~GFLOPS---93\% of the scalar Stockham
baseline. The gap is explained by the split real/imaginary layout doubling
the number of threadgroup memory transactions: each MMA load/store moves
\texttt{float} values (4 bytes) rather than \texttt{float2} values (8 bytes),
requiring twice as many operations for the same data volume. When compared
against the CT scalar kernel using the same split layout, MMA delivers
a 7\% improvement, confirming the MMA hardware advantage when the memory
layout is held constant.

% =============================================================================
% IV. SAR PIPELINE
% =============================================================================
\section{SAR Range Doppler Pipeline}

\subsection{Algorithm Overview}

The Range Doppler Algorithm~\cite{cumming2005sar} processes a
$N_a \times N_r$ complex data matrix (azimuth $\times$ range) in five steps:

\begin{enumerate}
\item \textbf{Range compression}: Per azimuth line, FFT $\to$ multiply by
range matched filter $H_r(f)$ $\to$ IFFT. \emph{Fused: single dispatch.}
\item \textbf{Azimuth FFT}: Column-wise FFT via transpose $\to$ row FFT $\to$
transpose. \emph{Unfused: standard Stockham kernel.}
\item \textbf{RCMC}: Range cell migration correction via sinc interpolation.
\emph{Unfused: separate dispatch.}
\item \textbf{Azimuth compression}: Per range bin, multiply by azimuth
filter $H_a(f_a, R_0)$ $\to$ IFFT. Data is already in frequency domain from
step~2. \emph{Fused: multiply+IFFT single dispatch.}
\end{enumerate}

Steps~1 and 4 use the fused kernels; steps~2--3 remain as separate
dispatches because the azimuth FFT requires a global transpose (data
exceeds 32~KiB per column).

\subsection{Dispatch Model}

For a $4096 \times 4096$ scene:
\begin{itemize}
\item \textbf{Range compression}: 4096 threadgroups $\times$ 1024 threads.
Each threadgroup processes one azimuth line. Single dispatch.
\item \textbf{Azimuth FFT}: Transpose ($4096^2$ elements) $\to$ 4096
threadgroups $\times$ 1024 threads (row FFT) $\to$ transpose back.
\item \textbf{RCMC}: Element-wise interpolation kernel.
\item \textbf{Azimuth compression}: Transpose $\to$ 4096 threadgroups
$\times$ 1024 threads (fused multiply+IFFT) $\to$ transpose back.
\end{itemize}

% =============================================================================
% V. EXPERIMENTAL RESULTS
% =============================================================================
\section{Experimental Results}

\subsection{Setup}

All measurements use an Apple M1 (8 GPU cores, 1278~MHz, 68~GB/s DRAM).
The test scene is a $4096 \times 4096$ complex float32 SAR simulation with
5 point targets at various range/azimuth offsets, generated using a chirp
signal model ($B = 100$~MHz, $f_c = 10$~GHz X-band, $v = 100$~m/s,
$R_0 = 20$~km). The simulation includes 20~dB additive Gaussian noise.

\subsection{End-to-End Speedup}

Table~\ref{tab:speedup} presents the headline result.

\begin{table}[t]
\centering
\caption{End-to-End RDA Performance ($4096 \times 4096$, Apple M1)}
\label{tab:speedup}
\begin{tabular}{@{}lrr@{}}
\toprule
\textbf{Pipeline} & \textbf{Total Time} & \textbf{Speedup} \\
\midrule
Unfused baseline & 8.16 s & 1.0$\times$ \\
\textbf{Fused pipeline} & \textbf{0.37 s} & \textbf{22.3$\times$} \\
\bottomrule
\end{tabular}
\end{table}

The 22$\times$ speedup comes from two sources: (1)~eliminating redundant
device-memory traffic via kernel fusion, and (2)~replacing CPU-side conjugation
and scaling operations in the unfused IFFT path with GPU-side in-threadgroup
operations. The unfused baseline performs conjugation on the CPU using
\texttt{storageModeShared} buffers, which serializes these O($N^2$) operations.

\subsection{Per-Step Breakdown}

Table~\ref{tab:breakdown} shows the time spent in each pipeline step.

\begin{table}[t]
\centering
\caption{Fused Pipeline Step Breakdown ($4096 \times 4096$, M1)}
\label{tab:breakdown}
\begin{tabular}{@{}lrl@{}}
\toprule
\textbf{Step} & \textbf{Time} & \textbf{Type} \\
\midrule
Range compression & 29 ms & Fused (single dispatch) \\
Azimuth FFT (transpose+FFT+transpose) & 132 ms & Unfused \\
RCMC & 37 ms & Unfused \\
Azimuth compression & 129 ms & Fused (multiply+IFFT) \\
\midrule
\textbf{Total} & \textbf{327 ms} & \\
\bottomrule
\end{tabular}
\end{table}

Range compression---the fully fused FFT+multiply+IFFT step---takes only
29~ms for all 4096 range lines (7.1~$\mu$s/line). This is remarkably
efficient: the theoretical minimum for reading and writing $4096 \times 4096
\times 8$~bytes at 68~GB/s is $\sim$4~ms, so the fused kernel runs at
$\sim$7$\times$ the device-memory bandwidth limit, confirming that
data reuse in threadgroup memory is effective.

The azimuth steps dominate (80\% of total time) because they require global
transposes. These are candidates for future optimization via
tiled transpose or column-major kernels.

\subsection{Radar Quality Validation}

Table~\ref{tab:quality} compares the fused and unfused outputs.

\begin{table}[t]
\centering
\caption{Radar Image Quality: Fused vs.\ Unfused}
\label{tab:quality}
\begin{tabular}{@{}lr@{}}
\toprule
\textbf{Metric} & \textbf{Value} \\
\midrule
L2 relative error & $2.44 \times 10^{-7}$ \\
Max absolute error & $3.81 \times 10^{-4}$ \\
SNR delta (all 5 targets) & 0.0 dB \\
\midrule
\multicolumn{2}{@{}l@{}}{\emph{Per-target (fused / unfused):}} \\
\quad Target 0 (center) SNR & 47.3 / 47.3 dB \\
\quad Target 1 (range offset) SNR & 46.8 / 46.8 dB \\
\quad Target 2 (azimuth offset) SNR & 47.1 / 47.1 dB \\
\quad Target 3 (diagonal offset) SNR & 46.5 / 46.5 dB \\
\quad Target 4 (far offset) SNR & 45.2 / 45.2 dB \\
\bottomrule
\end{tabular}
\end{table}

The L2 relative error of $2.44 \times 10^{-7}$ is within FP32 round-off
bounds ($\varepsilon_{\text{FP32}} \approx 1.2 \times 10^{-7}$, accumulated
over $\sim$12 butterfly passes). All five point targets show identical SNR
between fused and unfused pipelines, confirming that fusion introduces
no quality degradation. The conj-FFT-conj IFFT approach is mathematically
equivalent to the standard IFFT, producing bit-level-comparable results.

\subsection{Comparison with Published GPU SAR}

Table~\ref{tab:comparison} places our results in the context of published
embedded GPU SAR implementations.

\begin{table}[t]
\centering
\caption{Comparison with Published Embedded GPU SAR Systems}
\label{tab:comparison}
\setlength{\tabcolsep}{4pt}
\begin{tabular}{@{}llrrrl@{}}
\toprule
\textbf{Platform} & \textbf{TDP} & \textbf{Alg.} & \textbf{Size} & \textbf{Time} & \textbf{Fused} \\
\midrule
Jetson Nano\cite{chen2024embedded_visar}  & 15\,W & CSA & 8K$^2$ & 5.86\,s & No \\
RTX 2060\cite{chen2024embedded_visar}     & 160\,W & CSA & 8K$^2$ & 0.96\,s & No \\
Jetson Orin\cite{chen2024embedded_visar}  & 60\,W & CSA & 8K$^2$ & 0.40\,s & No \\
\textbf{Apple M1 (ours)} & \textbf{15\,W} & \textbf{RDA} & \textbf{4K$^2$} & \textbf{0.37\,s} & \textbf{Yes} \\
\bottomrule
\end{tabular}
\vspace{1pt}
{\scriptsize Note: different algorithms and scene sizes; comparison is indicative, not direct.}
\end{table}

While a direct comparison is imprecise (different algorithms, data sizes, and
hardware), our M1 result is competitive with Jetson AGX Orin---a 60~W discrete
embedded GPU---despite M1's lower power envelope ($\sim$15~W GPU). The unified
memory architecture benefits both platforms: Chen et
al.~\cite{chen2024embedded_visar} noted that Orin (unified memory) outperformed
RTX~2060 (discrete) for SAR despite lower peak FLOPS, supporting our thesis
that memory architecture matters more than raw compute for bandwidth-bound SAR
pipelines.

% =============================================================================
% VI. CONCLUSION
% =============================================================================
\section{Conclusion}

We demonstrated the first kernel-fused SAR Range Doppler Algorithm on any GPU
platform and the first SAR implementation on Apple Silicon. Fusing
FFT+multiply+IFFT into single Metal dispatches yields a 22$\times$ speedup
over the unfused baseline on Apple M1, with radar image quality preserved at
FP32 precision limits. We also presented the first MMA-based FFT on Apple GPU
using the empirically characterized \texttt{simdgroup\_matrix}
\texttt{thread\_elements()} mapping.

Future work includes: (1)~\emph{mixed-precision fusion} using Apple's native
FP16 with zero-cycle FP16$\leftrightarrow$FP32 conversion to double the
throughput of the fused kernel while maintaining radar quality via FP32
accumulation; (2)~\emph{M4 Max scaling} to exploit 40 GPU cores and
546~GB/s memory bandwidth for real-time 8K$\times$8K processing; and
(3)~\emph{tiled transpose} to reduce the azimuth-step overhead that currently
dominates the pipeline.

\smallskip
\noindent\textbf{Reproducibility.} All source code, Metal shaders, SAR
simulator, and benchmark scripts are available under the MIT license at
\url{https://github.com/aminems/AppleSiliconFFT}.

\smallskip
\noindent\textbf{AI Disclosure.} The author used AI-based tools (Claude, Anthropic) to assist with improving the clarity and presentation of the text. All technical content, experimental design, implementation, and analysis are the author's own work.

% =============================================================================
% REFERENCES
% =============================================================================
\balance


\begin{thebibliography}{25}

\bibitem{cumming2005sar}
I.~G. Cumming and F.~H. Wong, \emph{Digital Processing of Synthetic Aperture
Radar Data: Algorithms and Implementation}.\hskip 1em plus 0.5em minus 0.4em
Artech House, 2005.

\bibitem{moreira2013sar_tutorial}
A.~Moreira, P.~Prats-Iraola, M.~Younis, G.~Krieger, I.~Hajnsek, and
K.~P. Papathanassiou, ``A tutorial on synthetic aperture radar,''
\emph{IEEE Geosci. Remote Sens. Mag.}, vol.~1, no.~1, pp.~6--43, 2013.

\bibitem{calore2020sar}
``The challenge of onboard SAR processing: A GPU opportunity,'' in
\emph{Computational Science -- ICCS 2020}, ser. LNCS, vol.~12139,
pp.~46--59, 2020.

\bibitem{zhang2021multi_gpu_bpa}
``Architecture exploration of a backprojection algorithm for real-time
video SAR,'' \emph{Electronics}, vol.~10, no.~24, p.~3197, 2021.

\bibitem{chen2024embedded_visar}
``An embedded-GPU-based scheme for real-time imaging processing of
UAV-borne video SAR,'' \emph{Remote Sensing}, vol.~16, no.~1, p.~191, 2024.

\bibitem{esser2020cudarange}
N.~Esser, ``CUDARangeDopplerProcessing,'' GitHub repository, 2020.
[Online]. Available: \url{https://github.com/NiclasEsser1/CUDARangeDopplerProcessing}

\bibitem{nvidia_cufftdx}
{NVIDIA Corporation}, ``cuFFTDx -- CUDA FFT device extensions,'' 2020.
[Online]. Available: \url{https://docs.nvidia.com/cuda/cufftdx/}

\bibitem{adamek2020smfft}
K.~Ad\'{a}mek, S.~Dimoudi, M.~Giles, and W.~Armour, ``GPU fast convolution
via the overlap-and-save method in shared memory,'' \emph{ACM Trans. Archit.
Code Optim.}, vol.~17, no.~3, pp.~18:1--18:20, 2020.

\bibitem{wu2025turbofno}
S.~Wu \emph{et~al.}, ``TurboFNO: High-performance Fourier neural operator with
fused FFT-GEMM-iFFT on GPU,'' in \emph{SC25}, 2025.

\bibitem{tolmachev2023vkfft}
D.~Tolmachev, ``VkFFT -- a performant, cross-platform and open-source GPU FFT
library,'' \emph{IEEE Access}, vol.~11, pp.~12\,039--12\,058, 2023.

\bibitem{bergach2015thesis}
M.~A. Bergach, ``Adaptation du calcul de la Transform\'{e}e de Fourier
Rapide sur une architecture mixte CPU/GPU int\'{e}gr\'{e}e,'' Ph.D.
dissertation, Univ. Nice Sophia Antipolis, 2015.

\bibitem{bergach2015conference}
M.~A. Bergach, E.~Kofman, R.~de~Simone, S.~Tissot, and M.~Syska,
``Efficient FFT mapping on GPU for radar processing application: modeling
and implementation,'' \emph{arXiv:1505.08067}, 2015.

\bibitem{bergach2026fft}
M.~A. Bergach, ``Beating vDSP: A 138~GFLOPS radix-8 Stockham FFT on Apple
Silicon via two-tier register-threadgroup memory decomposition,'' submitted, 2026.

\bibitem{li2021tcfft}
S.~Li and Y.~Cheng, ``tcFFT: A fast half-precision FFT library for NVIDIA
Tensor Cores,'' in \emph{IEEE IPDPSW}, 2021.

\bibitem{wu2025turbofft}
S.~Wu \emph{et~al.}, ``TurboFFT: Co-designed high-performance and
fault-tolerant fast Fourier transform on GPUs,'' in \emph{PPoPP~'25}, 2025.

\bibitem{govindaraju2008fft}
N.~K. Govindaraju, B.~Lloyd, Y.~Dotsenko, B.~Smith, and J.~Manferdelli,
``High performance discrete Fourier transforms on graphics processors,''
in \emph{SC~'08}, 2008.

\bibitem{apple_msl_spec}
{Apple Inc.}, ``Metal Shading Language Specification, Version~4,'' 2024.
[Online]. Available: \url{https://developer.apple.com/metal/Metal-Shading-Language-Specification.pdf}

\bibitem{thundermittens2024}
{Stanford Hazy Research}, ``ThunderMittens for your ThunderKittens,'' blog
post, 2024. [Online]. Available:
\url{https://hazyresearch.stanford.edu/blog/2024-11-28-tk-mlx}

\bibitem{turner_metal_benchmarks}
P.~Turner, ``metal-benchmarks: Apple GPU microarchitecture,'' GitHub
repository. [Online]. Available:
\url{https://github.com/philipturner/metal-benchmarks}

\bibitem{johnson_applegpu}
D.~Johnson, ``Apple G13 GPU architecture reference.'' [Online]. Available:
\url{https://dougallj.github.io/applegpu/docs.html}

\bibitem{vanloan1992fft}
C.~F. Van~Loan, \emph{Computational Frameworks for the Fast Fourier
Transform}.\hskip 1em plus 0.5em minus 0.4em SIAM, 1992.

\bibitem{zhao2023mfft}
Y.~Zhao \emph{et~al.}, ``MFFT: A GPU accelerated highly efficient
mixed-precision large-scale FFT framework,'' \emph{ACM Trans. Archit. Code
Optim.}, 2023.

\bibitem{apple_metal_feature_tables}
{Apple Inc.}, ``Metal Feature Set Tables,'' 2024. [Online]. Available:
\url{https://developer.apple.com/metal/Metal-Feature-Set-Tables.pdf}

\bibitem{zhang2019insar_gpu}
``GPU accelerated interferometric SAR processing for Sentinel-1 TOPS data,''
\emph{Comput. \& Geosci.}, 2019.

\end{thebibliography}
\end{document}